\documentclass[aps,prl,twocolumn,superscriptaddress,showpacs,preprintnumbers,amsmath,amssymb]{revtex4}

\usepackage{graphicx} 
\usepackage{dcolumn} 
\usepackage{subfigure}
\usepackage[dvips]{color}

\newcommand{\gev} {\ensuremath{\, {\mathrm{GeV}}     }}
\newcommand{\mev} {\ensuremath{\, {\mathrm{MeV}}     }}
\newcommand{\gevc}{\ensuremath{\, {\mathrm{GeV}/c^2} }}
\newcommand{\mevc}{\ensuremath{\, {\mathrm{MeV}/c^2} }}

\newcommand{\ecm} {\ensuremath{ E_{\mathrm{c.m.}} }}
\newcommand{\gisr}{\ensuremath{ \gamma_{\mathrm{ISR}} }}
\newcommand{\sqs} {\ensuremath{ \sqrt{s} }}
\newcommand{\RM}  {\ensuremath{ M_{\mathrm{rec}}   }}

\newcommand{\ee}  {\ensuremath{ e^+e^- }}

\newcommand{\eell}     {\ensuremath{ e^+e^- \to \Lambda_c^+ \Lambda_c^-              }}
\newcommand{\eellg}    {\ensuremath{ e^+e^- \to \Lambda_c^+ \Lambda_c^- \gamma_{\mathrm{isr}} }}
\newcommand{\eellgmis} {\ensuremath{ e^+e^- \to \Lambda_c^+ \Lambda_c^- \pi^0        }}

\newcommand{\eellpg} {\ensuremath{ e^+e^- \to \Lambda_c^+ \Lambda_c^- \pi^0  \gamma_{\mathrm{isr}}  }}
\newcommand{\eellppg}{\ensuremath{ e^+e^- \to \Lambda_c^+ \Lambda_c^- \pi \pi \gamma_{\mathrm{isr}} }}
\newcommand{\eelsg}  {\ensuremath{ e^+e^- \to \Lambda_c^+ \Sigma_c^-  \gamma_{\mathrm{isr}}  }}

\newcommand{\lala} {\ensuremath{ \Lambda_c^+  \Lambda_c^-         }}
\newcommand{\llp}  {\ensuremath{ \Lambda_c^+  \Lambda_c^- \pi^0   }}
\newcommand{\llpp} {\ensuremath{ \Lambda_c^+  \Lambda_c^- \pi \pi }}

\newcommand{\ls}   {\ensuremath{ \Lambda_c^+  \Sigma_c^-         }}
\newcommand{\llfi} {\ensuremath{ \Lambda_c^+  \Lambda_c^-(2595)  }}
\newcommand{\lls}  {\ensuremath{ \Lambda_c^+  \Lambda_c^-(2625)  }}
\newcommand{\llt}  {\ensuremath{ \Lambda_c^+  \Lambda_c^-(2765)  }}
\newcommand{\llfo} {\ensuremath{ \Lambda_c^+  \Lambda_c^-(2880)  }}

\newcommand{\la}   {\ensuremath{ \Lambda_c   }}
\newcommand{\lap}  {\ensuremath{ \Lambda_c^+ }}
\newcommand{\lam}  {\ensuremath{ \Lambda_c^- }}
\newcommand{\sm}   {\ensuremath{ \Sigma_c^- }}

\newcommand{\mll}  {\ensuremath{ M_{\Lambda_c^+  \Lambda_c^-  }}} 
\newcommand{\mlap} {\ensuremath{ M_{\Lambda_c^+ }}}

\newcommand{\eeddp}    {\ensuremath{ e^+e^- \to D^0 D^- \pi^+ }}
\newcommand{\psiddt}   {\ensuremath{ \psi(4415)\to D \overline D{}^{*}_2(2460) }}
\newcommand{\eedd}     {\ensuremath{ e^+e^- \to D   \overline D     }}

\newcommand{\eeddst}   {\ensuremath{ e^+e^- \to D^{(*)\pm}{D}{}^{*\mp} }}

\begin{document}

\title{\quad\\[0.5cm] Observation of a near-threshold enhancement in
  the \eell\ cross section using initial-state radiation}

\affiliation{Budker Institute of Nuclear Physics, Novosibirsk}
\affiliation{Chiba University, Chiba}
\affiliation{University of Cincinnati, Cincinnati, Ohio 45221}
\affiliation{T. Ko\'{s}ciuszko Cracow University of Technology, Krakow}
\affiliation{Justus-Liebig-Universit\"at Gie\ss{}en, Gie\ss{}en}
\affiliation{The Graduate University for Advanced Studies, Hayama}
\affiliation{Gyeongsang National University, Chinju}
\affiliation{Hanyang University, Seoul}
\affiliation{University of Hawaii, Honolulu, Hawaii 96822}
\affiliation{High Energy Accelerator Research Organization (KEK), Tsukuba}
\affiliation{Institute of High Energy Physics, Chinese Academy of Sciences, Beijing}
\affiliation{Institute of High Energy Physics, Vienna}
\affiliation{Institute of High Energy Physics, Protvino}
\affiliation{Institute for Theoretical and Experimental Physics, Moscow}
\affiliation{J. Stefan Institute, Ljubljana}
\affiliation{Kanagawa University, Yokohama}
\affiliation{Korea University, Seoul}
\affiliation{Kyungpook National University, Taegu}
\affiliation{\'Ecole Polytechnique F\'ed\'erale de Lausanne (EPFL),
Lausanne}
\affiliation{Faculty of Mathematics and Physics, University of Ljubljana,
Ljubljana}
\affiliation{University of Maribor, Maribor}
\affiliation{University of Melbourne, School of Physics, Victoria 3010}
\affiliation{Nagoya University, Nagoya}
\affiliation{Nara Women's University, Nara}
\affiliation{National Central University, Chung-li}
\affiliation{National United University, Miao Li}
\affiliation{Department of Physics, National Taiwan University, Taipei}
\affiliation{H. Niewodniczanski Institute of Nuclear Physics, Krakow}
\affiliation{Nippon Dental University, Niigata}
\affiliation{Niigata University, Niigata}
\affiliation{University of Nova Gorica, Nova Gorica}
\affiliation{Osaka City University, Osaka}
\affiliation{Osaka University, Osaka}
\affiliation{Panjab University, Chandigarh}
\affiliation{Saga University, Saga}
\affiliation{University of Science and Technology of China, Hefei}
\affiliation{Seoul National University, Seoul}
\affiliation{Sungkyunkwan University, Suwon}
\affiliation{University of Sydney, Sydney, New South Wales}
\affiliation{Toho University, Funabashi}
\affiliation{Tohoku Gakuin University, Tagajo}
\affiliation{Department of Physics, University of Tokyo, Tokyo}
\affiliation{Tokyo Institute of Technology, Tokyo}
\affiliation{Tokyo Metropolitan University, Tokyo}
\affiliation{Tokyo University of Agriculture and Technology, Tokyo}
\affiliation{Virginia Polytechnic Institute and State University,
Blacksburg, Virginia 24061}
\affiliation{Yonsei University, Seoul}

\author{G.~Pakhlova}\affiliation{Institute for Theoretical and  Experimental Physics, Moscow} 
\author{I.~Adachi}\affiliation{High Energy Accelerator Research Organization (KEK), Tsukuba} 
\author{H.~Aihara}\affiliation{Department of Physics, University of Tokyo, Tokyo} 
\author{K.~Arinstein}\affiliation{Budker Institute of Nuclear Physics, Novosibirsk} 
\author{V.~Aulchenko}\affiliation{Budker Institute of Nuclear Physics, Novosibirsk} 
\author{T.~Aushev}\affiliation{\'Ecole Polytechnique F\'ed\'erale de Lausanne (EPFL), Lausanne}
\affiliation{Institute for Theoretical and Experimental Physics, Moscow} 
\author{A.~M.~Bakich}\affiliation{University of Sydney, Sydney, New South Wales} 
\author{V.~Balagura}\affiliation{Institute for Theoretical and Experimental Physics, Moscow} 
\author{I.~Bedny}\affiliation{Budker Institute of Nuclear Physics, Novosibirsk} 
\author{V.~Bhardwaj}\affiliation{Panjab University, Chandigarh} 
% ...

\collaboration{The Belle Collaboration}

\begin{abstract}

We report a measurement of the exclusive \eell\ cross section as a
function of center-of-mass energy near the \lala\ threshold.  A clear
peak with a significance of $8.2\sigma$ is observed in the
\lala\ invariant mass distribution just above threshold. With an
assumption of a resonance origin for the observed peak, a mass and
width of $M=(4634^{+8}_{-7} \mathrm{(stat.)}  ^{+5}_{-8}
\mathrm{(sys.)})\mevc$ and $\Gamma_{\mathrm{tot}}=(92^{+40}_{-24}
\mathrm{(stat.)}^{+10}_{-21} \mathrm{(sys.)})\mev$ are determined.
The analysis is based on a study of events with
initial-state-radiation photons in a data sample collected with the
Belle detector at the $\Upsilon(4S)$ resonance and nearby continuum
with an integrated luminosity of $695$ $\mathrm{fb}^{-1}$ at the KEKB
asymmetric-energy \ee\ collider.

\end{abstract}


\maketitle
\setcounter{footnote}{0}

The discovery of many unexpected charmonium-like states has stimulated
renewed interest in charmonium physics. Among these new states, the
$Y(4260)$~\cite{babar:4260,belle:4260}, $Y(4360)$ and
$Y(4660)$~\cite{babar:4360,belle:4360} have quantum numbers
$J^{PC}=1^{--}$ and are produced via \ee\ annihilation. Surprisingly,
no evidence for open-charm production associated with these new states
has been observed.  Moreover, the parameters of the conventional
charmonium $1^{--}$ states obtained from fits to the inclusive cross
section~\cite{bes:fit} remain poorly understood
theoretically~\cite{barnes}. Measurements of exclusive cross sections
for charmed meson and baryon pairs in the 4 to 5 \gev\ energy range
are needed to help to clarify the situation.

Initial-state radiation (ISR) provides a powerful tool for measuring
exclusive \ee\ cross sections at \sqs\ smaller than the initial
\ee\ center-of-mass (c.m.) energy (\ecm) at $B$-factories. ISR allows
one to obtain cross sections over a broad energy range, while the high
luminosity of the $B$-factories compensates for the suppression
associated with the emission of a hard photon.  The first measurements
of the exclusive cross sections for \eeddst\ for \sqs\ near the
$D^{(*)\pm}D^{*\mp}$ thresholds were performed by
Belle~\cite{belle:ddst}. Subsequently, BaBar~\cite{babar:dd} and
Belle~\cite{belle:dd} presented exclusive \eedd\ production
measurements via ISR. Recently, Belle~\cite{belle:dd2} reported a
measurement of the exclusive cross section for \eeddp\ ~\cite{foot}
and the first observation of \psiddt\ decay.  These measured final
states almost saturate the total cross section for hadron production
in \ee\ annihilation in the \sqs\ region up to $\sim 4.3\gev$.  The
thresholds for charm baryon-antibaryon pair production lie in the
energy range above $4.5\gev$, where experimental data are
limited~\cite{lamX} or unavailable.

In this Letter we report the first measurement of the exclusive cross
section for the process \eell\ via ISR and the first observation of a
resonant-like structure at threshold.  The data sample corresponds to
an integrated luminosity of $695\,\mathrm{fb}^{-1}$ collected with the
Belle detector~\cite{det} at the $\Upsilon(4S)$ resonance and nearby
continuum at the KEKB asymmetric-energy \ee\ collider~\cite{kekb}.

The selection of \eellg\ signal events using full reconstruction of
both the \lap\ and \lam\ baryons suffers from the low
\la\ reconstruction efficiency and small branching fractions for
decays to accessible final states. Therefore, in order to achieve
higher efficiency we require full reconstruction of only one of the
\la\ baryons and the \gisr\ photon.  In this case the spectrum of
masses recoiling against the $\lap\gisr$ system,
\begin{eqnarray}
\RM(\lap\gisr)\!=\!\sqrt{(\ecm\!-\!E{}^*_{\lap\gisr}){}^2\!-\!p{}^{*2}_{\lap\gisr}},
\end{eqnarray}
peaks at the \lam\ mass. Here $E{}^*_{\lap\gisr}$ and
$p{}^*_{\lap\gisr}$ are the center-of-mass energy and momentum,
respectively, of the $\lap\gisr$ combination. The $\RM(\lap\gisr)$
peak is broad ($\sigma_{\RM} \sim 250\mevc$ according to a Monte Carlo
(MC) simulation) and asymmetric due to the photon energy resolution
and higher-order ISR processes ({\it i.e.} more than one \gisr\ in the
event). This makes the distinction between \lala, \llp\ and
\llpp\ final states difficult.

For the measurement of the exclusive cross section for \eell, we
determine the mass recoiling against the \gisr\ photon ($\RM(\gisr))$,
which is equivalent to $M(\lala)$ in the absence of higher-order QED
processes. To improve the $\RM(\gisr)$ resolution (expected to be
$\sim 100\mevc$), we apply a refit that constrains $\RM(\lap\gisr)$ to
the nominal \lam\ mass. In this way we use the well measured
properties of the fully reconstructed \lap\ to correct the poorly
measured energy of the \gisr.  As a result, the $M_{\lala}$ resolution
is improved substantially; it varies from $\sim 3\mevc$ just above
threshold to $\sim 8 \mevc$ at $\mll\sim 5.4 \gevc$.

All charged tracks are required to originate from the vicinity of the
interaction point (IP); we impose the requirements $dr<1 \,
{\mathrm{cm}}$ and $|dz|<4\,{\mathrm{cm}}$, where $dr$ and $|dz|$ are
the impact parameters perpendicular to and along the beam direction
with respect to the IP.  Particle identification requirements are
based on {\it dE/dx}, aerogel Cherenkov and time-of-flight counter
information~\cite{nim}.  Protons and charged kaons have typical
misidentification probabilities less than 0.1.  No identification
requirements are applied for pion candidates.  $K^0_S$ ($\Lambda$)
candidates are reconstructed from $\pi^+ \pi^-$($p \pi^-$) pairs with
an invariant mass within $10\mevc$ ($\sim 3\sigma$) of the $K^0_S$
($\Lambda$) mass. The distance between the two pion (proton and pion)
tracks at the $K^0_S$ ($\Lambda$) vertex must be less than
$1\,\mathrm{cm}$, the transverse flight distance from the interaction
point is required to be greater than $0.1\,\mathrm{cm}$, and the angle
between the $K^0_S$($\Lambda$) momentum direction and the flight
direction in the $x-y$ plane should be less than $0.01
(0.005)\,\mathrm{rad}$.  Photons are reconstructed in the
electromagnetic calorimeter as showers with energies greater than $50
\mev$ that are not associated with charged tracks. ISR photon
candidates are required to have energies greater than $3.5 \gev$.
Candidate $\pi^0$ mesons are formed from pairs of photons.  If the
mass of a $\gamma \gamma$ pair lies within $15\mevc$ ($\sim 3\sigma$)
of the $\pi^0$ mass, the pair is fit with a $\pi^0$ mass constraint
and considered as a $\pi^0$ candidate.

\begin{figure}[htb]
\begin{tabular}{cc}
\hspace*{-0.025\textwidth}
\includegraphics[width=0.48\textwidth]{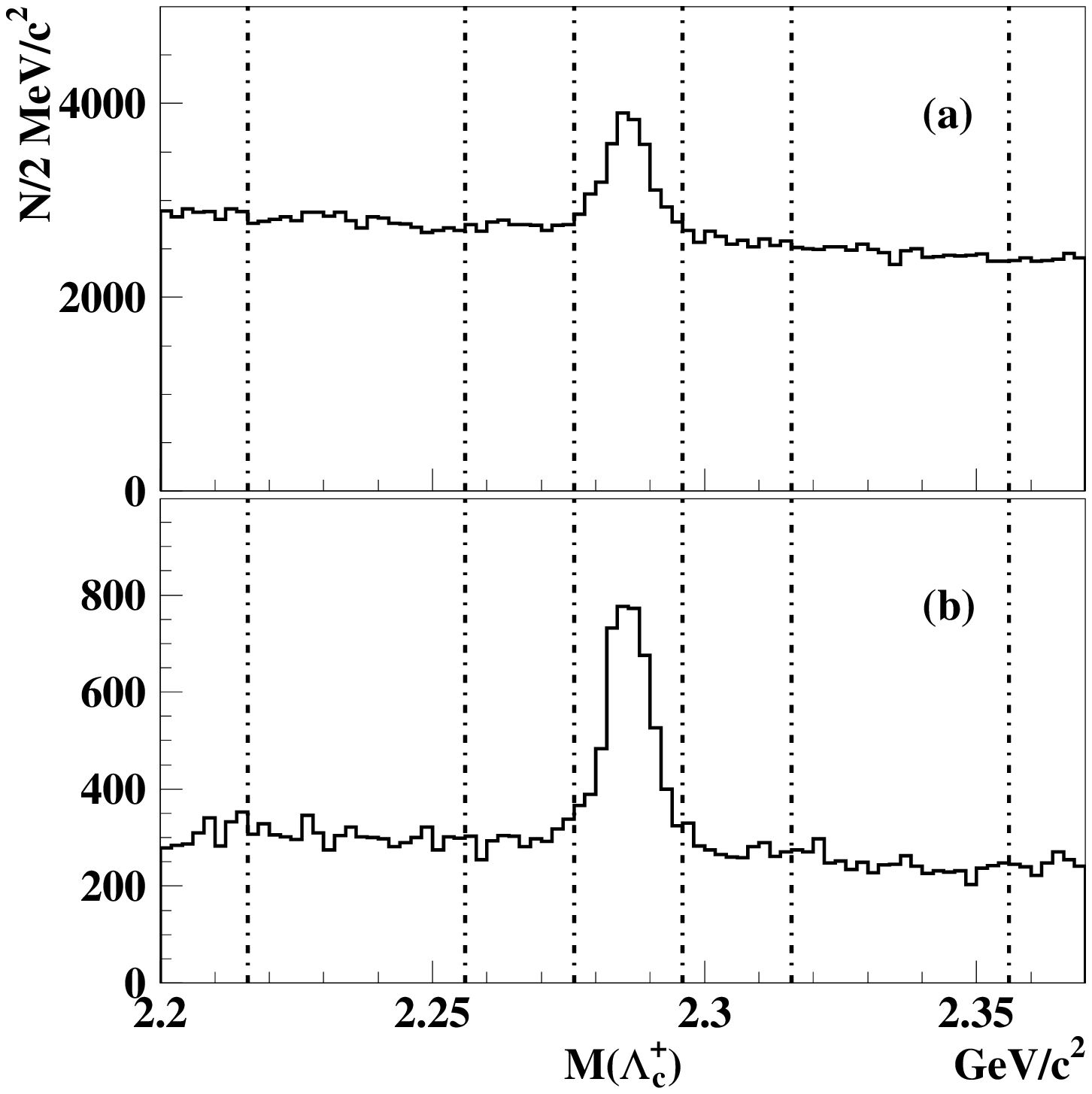}
\end{tabular}
\caption{The mass distribution of \lap\ from $\lap\gisr$ combinations:
  (a) without a $\overline p$ tag; (b) with a $\overline p$ tag.  The
  \lap\ signal and background regions are indicated by vertical
  lines.}
\label{Fig1}
\end{figure}

\lap\ candidates are reconstructed using three decay modes: $p K^0_S$,
$p K^-\pi^+$ and $\Lambda \pi^+$.  The mass distribution of
\lap\ candidates from $\lap\gisr$ combinations is shown in
Fig.~\ref{Fig1}\,(a). To suppress combinatorial background, we require
the presence of at least one $\overline p$ in the event from the decay
of the unreconstructed \lam\ ($\overline p$ tag).  As a result, the
combinatorial background is suppressed by a factor of $\sim 10$ at the
expense of about a $40\%$ reduction in signal according to the MC
simulation (see Fig.~\ref{Fig1}\,(b)).

A $\pm 10\mevc$ mass window is used for all \lap\ candidate decay
modes ($\sim 2.5\,\sigma$ in each case). To improve the momentum
resolution of \lap\ candidates, final tracks are fitted to a common
vertex with a mass constraint to the \lap\ mass. Only one $\lap\gisr$
combination per event is accepted; in the case of multiple
combinations, which occur in $5\%$ of the candidate events, the
combination with the best $\chi^2$ for the \lap\ mass fit is
selected. \lap\ mass sidebands selected for the background study are
four times as large as the signal region.  To avoid signal
over-subtraction, the sidebands are shifted by $20\mevc$ from the
signal region.  The sidebands are divided into windows of the same
width as that for the signal.  The \lap\ candidates from these
sidebands are refitted to the central mass value of each window and a
single candidate in each window per event is selected.

\begin{figure}[htb]
\begin{tabular}{cc}
\hspace*{-0.025\textwidth}
\includegraphics[width=0.48\textwidth]{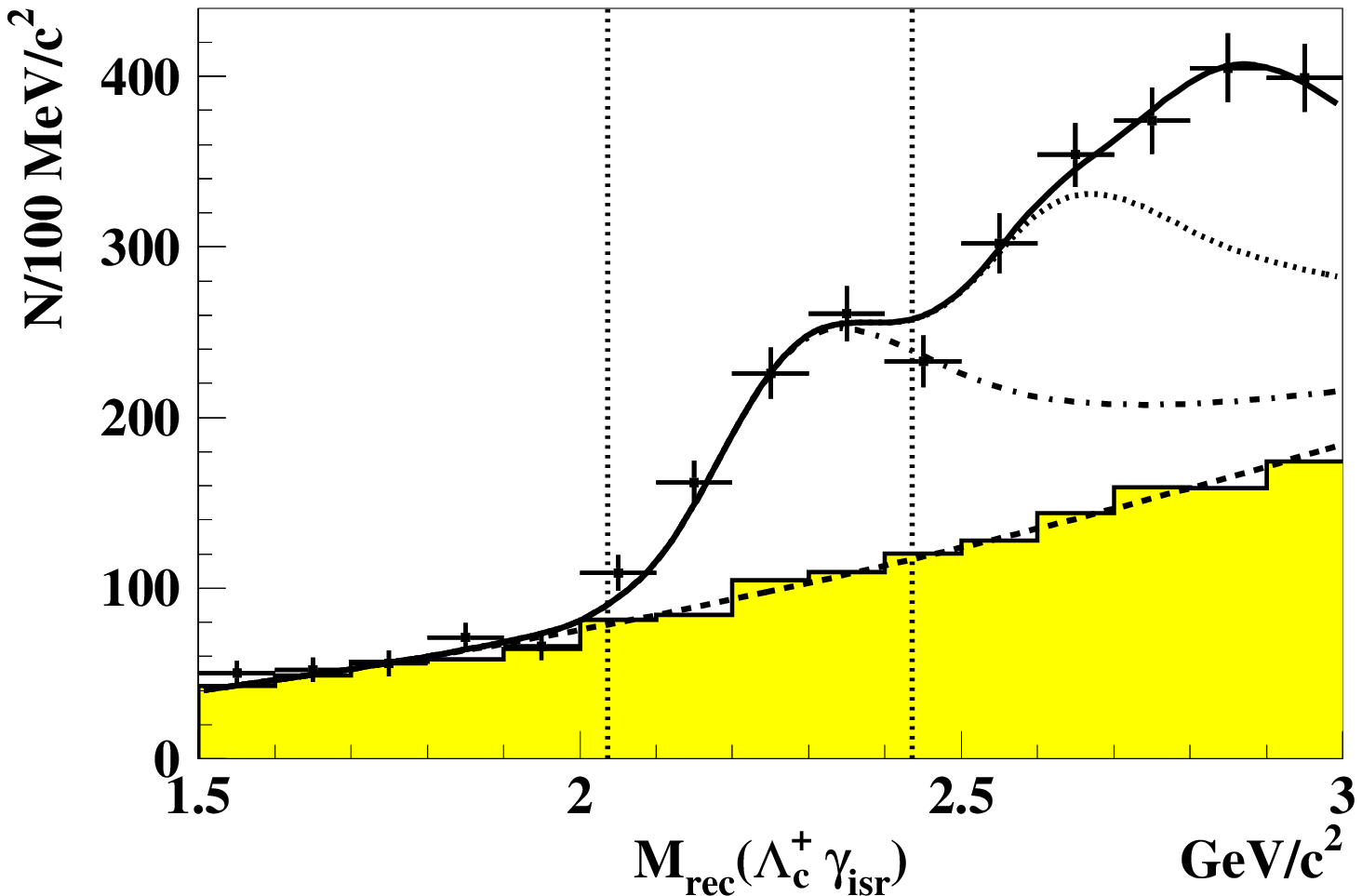}
\end{tabular}
\caption{The $\RM (\lap\gisr)$ distribution with a $\overline p$
  tag. The solid curve represents the result of the fit described in
  the text.  The combinatorial background parameterization is shown by
  the dashed curve. The dashed-dotted curve represents a contribution
  of the \lala\ final state while the dotted one is that of the
  \llfi\ and the \lls\ final states. The difference between the solid
  and dotted lines corresponds to the contribution of the \llt\ and
  the \llfo\ final states.  The histogram shows the normalized
  $M_{\lap}$ sidebands contributions. The selected signal window is
  indicated by the vertical lines. }
\label{Fig2}
\end{figure}

The distribution of $\RM(\lap\gisr)$ with a $\overline p$ tag is shown
in Fig.~\ref{Fig2}.
The excess around the \lam\ mass includes the \eellg\ signal as well
as possible reflections from the \eellpg\ and \eellppg\ processes with
an additional $\pi^0$ or $\pi\pi$, respectively, in the final
state. The process \eellpg, which could proceed via \eelsg, violates
isospin and is expected to be strongly suppressed. The process
\eellppg\ is allowed and is expected to proceed via \llfi, \lls,
\llt\ and \llfo\ final states.  Each final state would produce a broad
peak in the $\RM(\lap\gisr)$ distribution around the corresponding
mass value ({\it i.e.} $m_{\sm}$, $m_{\Lambda^{-}(2595)}$,
$m_{\Lambda^{-}(2625)}$, $m_{\Lambda^{-}(2765)}$ {\rm and}
$m_{\Lambda^{-}(2880)}$). Due to the poor $\RM(\lap\gisr)$ resolution
these peaks overlap and appear as a shoulder for masses above $\sim
2.5\gevc$.

To estimate the contribution from the reflections and to optimize the
signal region requirement, we fit the $\RM(\lap\gisr)$ distribution
with the sum of a signal plus a combinatorial and reflection
background with normalizations left as free parameters. To describe
the combinatorial background, we use \lap\ sideband data parameterized
by a second-order polynomial. We perform a simultaneous likelihood fit
to the $\RM(\lap\gisr)$ signal and sideband spectra. The signal and
reflection shapes of the \ls, \llfi, \lls, \llt, \llfo\ final states
are fixed from the MC simulation.  All reflection normalizations are
floated separately in the fit.  The goodness of the fit is found to be
$\chi^2/n.d.f=18.8/22$. We define an asymmetric requirement on
$\RM(\lap \gisr)$ of $-250 \mevc < m_{\lam} < 150 \mevc$ to suppress
the dominant part of the reflection background, as shown in
Fig.~\ref{Fig2}. We find $(386 \pm 27\mathrm{(stat.)})$ signal events
in this signal region. The contribution of the process \eellpg\ in the
signal region is estimated to be less than 18 events at the $90\%$
C.L. while that from the \eellppg\ process is estimated to be $(7.3
\pm 1.7 \mathrm{(stat.)})$ events.  In the following study the
possible contribution of these backgrounds is included in the
systematic error.

The contribution from \eellgmis , where an energetic $\pi^0$ is
misidentified as a single \gisr, is found to be negligibly small.
This is determined from a study of \eellgmis\ events using a similar
reconstruction technique, but with an energetic $\pi^0$ replacing the
\gisr.

\begin{figure}[htb]
\begin{tabular}{cc}
\hspace*{-0.025\textwidth}
\includegraphics[width=0.47\textwidth]{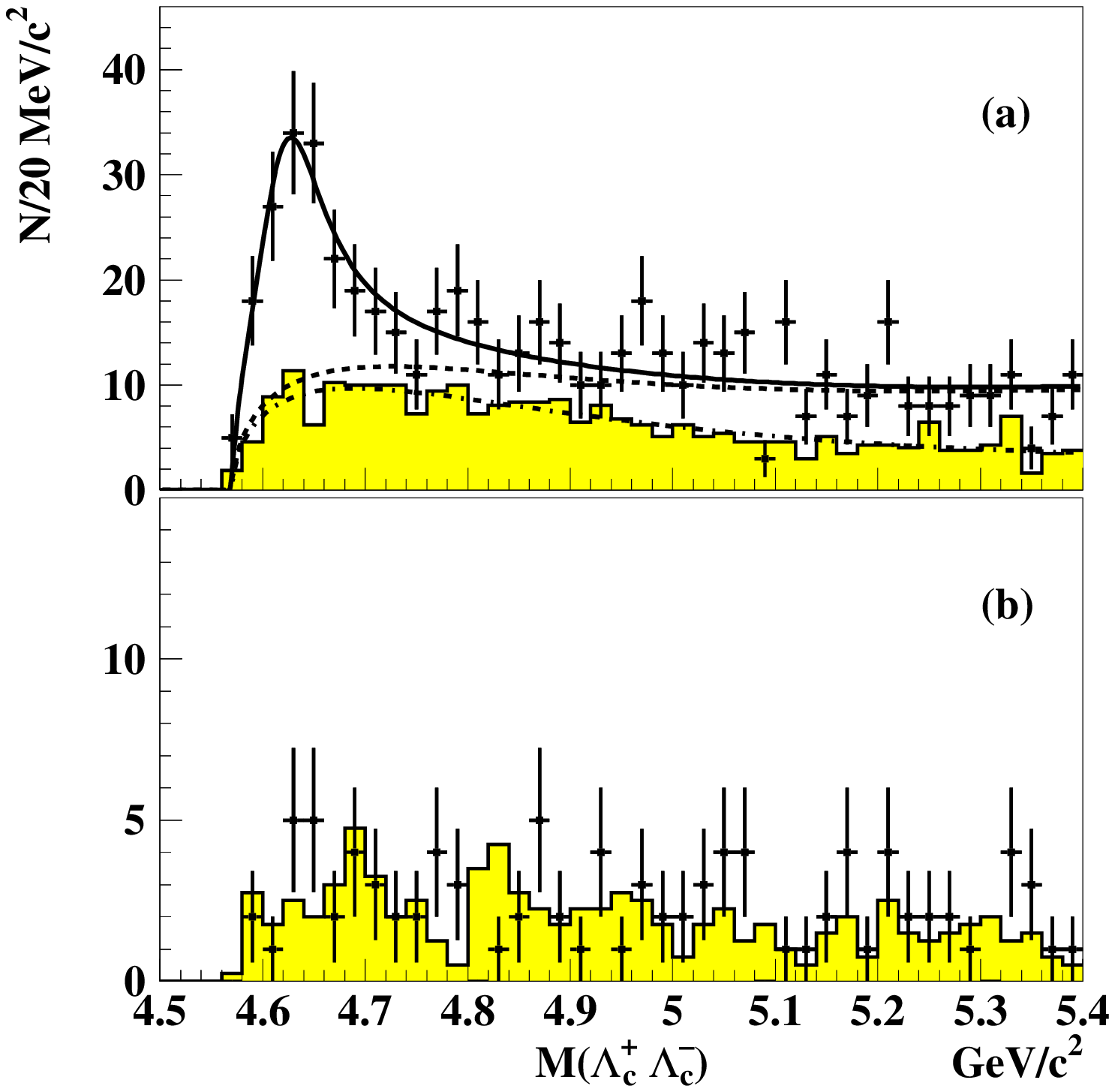}
\end{tabular}
\caption{The \mll\ spectrum for the signal region: (a) with $\overline
  p$ tag. The solid curve represents the result of the fit described
  in the text. The threshold function is shown by the dashed
  curve. The combinatorial background parameterization is shown by the
  dashed-dotted curve; (b) with proton (wrong-sign) tag. Histograms
  show the normalized contributions from \lap\ sidebands.}
\label{Fig3}
\end{figure}

The \mll\ spectrum for events in the signal region is shown in
Fig.~\ref{Fig3}\,(a).
A clear peak is evident near the \lala\ threshold.  We perform a
simultaneous likelihood fit to the \mll\ distributions for the
\lap\ signal and sideband regions to fix the combinatorial background
shapes. The combinatorial background is parameterized by
$p_1\sqrt{M-M_{\text{thr}}}\cdot e^{-(p_2\cdot M + p_3\cdot M^2)}$,
where $p_1$, $p_2$ and $p_3$ are free parameters.  The signal function
is a sum of a relativistic {\it s}-wave Breit-Wigner (RBW)
function~\cite{pdg} and a threshold function $\sqrt{M-M_{\text{thr}}}$
with a floating normalization to take into account a possible
non-resonant contribution.  Finally, the sum of the signal resonance
and non-resonant functions is multiplied by an efficiency function
that has a linear dependence on \mll, and the differential ISR
luminosity, described in Ref.~\cite{belle:ddst}.  The fit, shown as a
solid curve in Fig.~\ref{Fig3}\,(a), attributes $142^{+32}_{-28}
\mathrm{ (stat.) }$ events to the RBW signal.  The obtained peak mass
is $M=(4634^{+8}_{-7} \mathrm{(stat.)}
^{+5}_{-8}\mathrm{(sys.)})\mevc$ and the total width is
$\Gamma_{\mathrm{tot}}=(92^{+40}_{-24}\mathrm{(stat.)}^{+10}_{-21}
\mathrm{(sys.)})\mev$. The fit gives $\chi^2/n.d.f=104/77$. Here the
systematic uncertainties are obtained by varying the fit range,
histogram bin size, efficiency function, parameterization of the
background function and the non-resonant parametrization. The
systematic error associated with the possible interference between the
resonance and non-resonant contributions is estimated from the fit
with a coherent sum of the RBW and non-resonant amplitudes, which has
the quality $\chi^2/n.d.f=103/76$ and yields a smaller mass
($4626\mevc$) and total width ($77\mev$). A statistical significance
for the signal of $8.8 \sigma$ is determined from the quantity
$-2\ln(\mathcal{L}_0 / \mathcal{L}_{\text{max}})$, where
$\mathcal{L}_{\text{max}}$ is the maximum likelihood returned by the
fit, and $\mathcal{L}_0$ is the likelihood with the amplitude of the
Breit-Wigner function set to zero, taking the reduction in the number
of degrees of freedom into account. The significance including
systematics is $8.2\sigma$.  We use X(4630) to denote the observed
structure.

As a cross check, we present in Fig.~\ref{Fig3}\,(b) the
\mll\ spectrum for the signal region for wrong-sign tags, {\em i.e.}
requiring a presence of a proton in the event in addition to the
$\lap\gisr$ combination.  The \mll\ distribution from the signal
\lap\ window is in good agreement with the normalized contributions
from the \lap\ sidebands.

\begin{figure}[htb]
\begin{tabular}{cc}
\hspace*{-0.025\textwidth}
\includegraphics[width=0.47\textwidth]{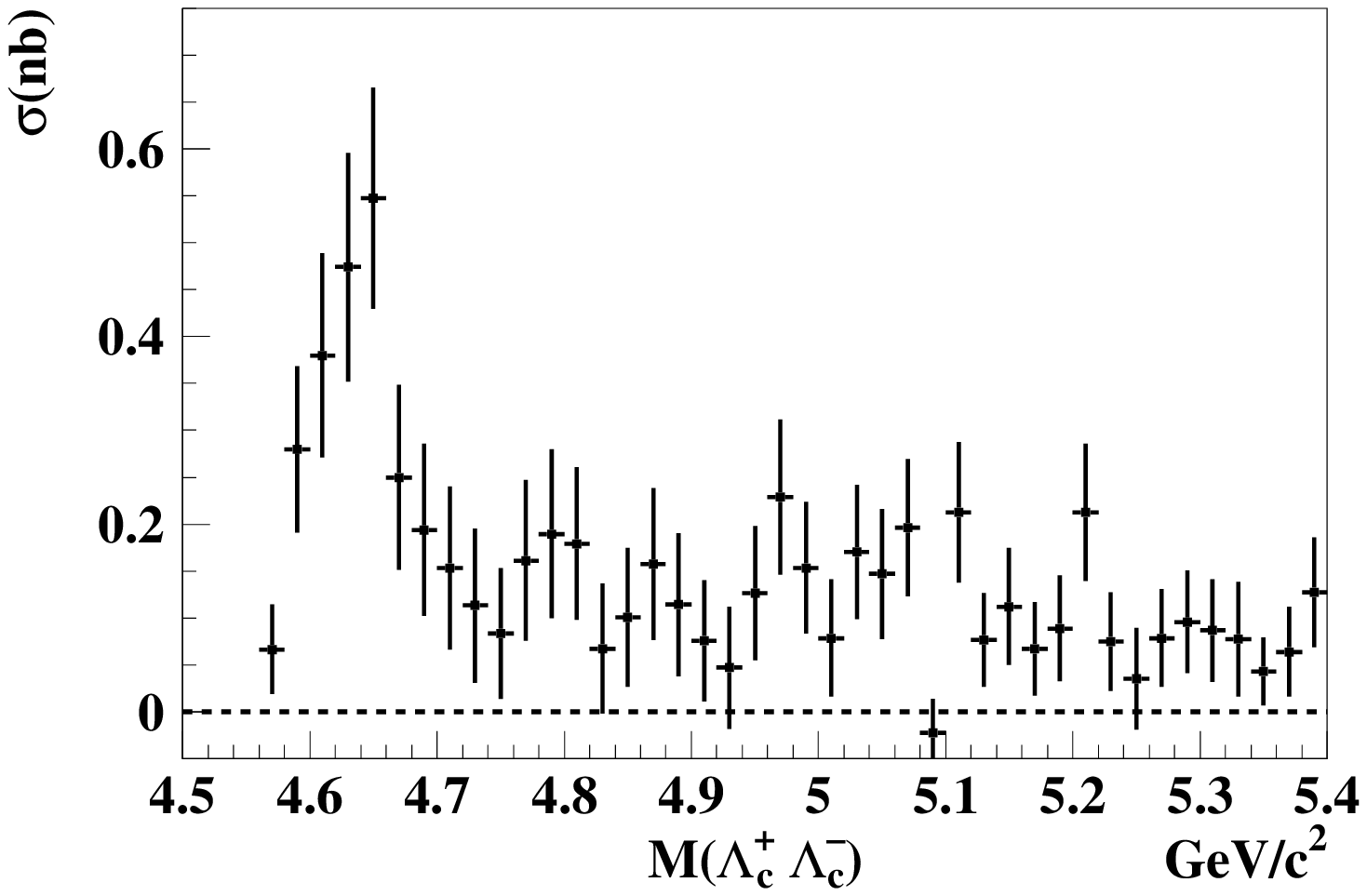} &
\end{tabular}
\caption{The cross section for the exclusive process \eell.}
\label{Fig4}
\end{figure}

The \eell\ cross section is extracted from the background-subtracted
\lala\ mass distribution following the procedure described in
Ref.~\cite{belle:ddst}, taking into account the differential ISR
luminosity and the efficiency function. The resulting \eell\ exclusive
cross section is shown in Fig.~\ref{Fig4} with statistical
uncertainties only. Since the bin width is much larger than
resolution, no correction for resolution is applied.

The peak cross section for the \eell\ process at $\ecm = m_{X(4630)}$
is calculated from the amplitude of the RBW function in the fit to be
$\sigma( \ee \to X(4630)) \times \mathcal{B}( X(4630) \to \lala) =
(0.47^{+0.11}_{-0.10} \mathrm{(stat.)}^{+0.05}_{-0.08}
\mathrm{(sys.)}\pm 0.19 \mathrm{(sys.))}$ nb. Here the first
systematic uncertainty is obtained by varying the fit range, histogram
bin, parameterization of the background function, efficiency and the
possible interference between the resonance and non-resonant
contributions. The second one comes from the uncertainties in
$\mathcal{B}( \lap \to p K^- \pi^+) = (5.0 \pm 1.3)\cdot 10^{-2}$ and
$\mathcal{B}(\lam\to\overline p X)= (50 \pm 16)\cdot
10^{-2}$~\cite{pdg}.  Using $\sigma(\ee \to X(4630)) = 12\pi /
m^2_{X(4630)} \times (\Gamma_{ee}/ \Gamma_{\mathrm{tot}})$ and the
$X(4630)$ mass value obtained from the fit we calculate $\Gamma_{ee}/
\Gamma_{\mathrm{tot}} \times \mathcal{B}(X(4630) \to \lala) =
(0.68^{+0.16}_{-0.15}\mathrm{(stat)}^{+0.07}_{-0.11} \mathrm{(sys.)}
\pm 0.28 \mathrm{(sys.))}\cdot 10^{-6}$.

The various contributions to the systematic errors for the $\sigma(
\ee \to \lala)$ measurements are summarized in Table ~\ref{tab1}.
\begin{table}[htb]
\caption{Contributions to the systematic error on the cross sections,
  [$\%$].}
\label{tab1}
\begin{center}
\begin{tabular}{@{\hspace{0.4cm}}l@{\hspace{0.4cm}}||@{\hspace{0.4cm}}c@{\hspace{0.4cm}}}
\hline \hline
Source & \lala\
\\ \hline
Background subtraction    & $\pm 7$ \\
Cross section calculation & $\pm 3$ \\
Reconstruction            & $\pm 5$ \\
Identification            & $\pm 3$ \\
Angular distributions     & $\pm 4$ \\
\hline
Total & $\pm 10$ \\
\hline 
$\mathcal{B}(\lap)$       & $\pm 26$ \\
$\mathcal{B}(\lam\to\overline p X)$ & $\pm 32$ \\
\hline \hline
\end{tabular}
\end{center}
\end{table}
The systematic errors associated with the combinatorial background
subtraction are estimated to be 3\% due to an uncertainty in the
scaling factors for the sideband subtractions. It is estimated using
fits to the \mlap\ distribution with different signal and background
parameterizations. Reflections from the \eellpg\ and
\eellppg\ processes are estimated conservatively to be smaller than
6\% of the signal. The uncertainty due to a possible
\eellgmis\ contribution is found to be 1\%.  The systematic error
ascribed to the cross section calculation includes a 1.5\% error on
the differential luminosity and 2\% error due to the MC
statistics. Another source of systematic error comes from
uncertainties in track and photon reconstruction efficiencies (1\% per
track and 1.5\% per photon). Another contribution comes from the
uncertainty in the kaon and proton identification efficiency. The
systematic uncertainty due to the unknown helicity angle distribution
for the \lala\ final state is included.  For the efficiency
calculation we use a flat helicity distribution and consider the
extreme cases $\mathrm{dN/d cos\theta\sim 1 + cos^2\theta}$ and
$\mathrm{\sim sin^2\theta}$ for the efficiency uncertainty.

In summary, we report the first measurements of the \eell\ exclusive
cross section over the center-of-mass energy range from the threshold
to 5.4\gev\ with initial-state radiation. We observe a significant
near-threshold enhancement in the studied cross section. The nature of
this enhancement remains unclear.  In many processes including
three-body $B$ meson baryonic decays, mass peaks are observed near
threshold~\cite{dibaryon}. However, the cross section for $\ee \to
\Lambda \overline \Lambda$ measured via ISR by BaBar~\cite{babar:lam}
has a different pattern: it increases sharply at threshold and then
decreases gradually without any peak-like structure. Assuming the
observed peak to be a resonance, its mass and width are found to be $M
=(4634^{+8}_{-7} \mathrm{(stat.)}  ^{+5}_{-8}\mathrm{(sys.)})\mevc$
and $\Gamma_{\mathrm{tot}}=(92^{+40}_{-24} \mathrm{(stat.)}
^{+10}_{-21} \mathrm{(sys.)})\mev$, respectively. These values are
consistent within errors with the mass and width of a new $1^{--}$
charmonium-like state, the Y(4660), that was found in $\psi(2S)\pi\pi$
decays via ISR~\cite{belle:4360}. Finally, we cannot exclude the
possibility that the observed enhancement is the $5^3S_1$ charmonium
state that is predicted around the observed mass ~\cite{badalyan}.

We thank the KEKB group for excellent operation of the accelerator,
the KEK cryogenics group for efficient solenoid operations, and the
KEK computer group and the NII for valuable computing and SINET3
network support.  We acknowledge support from MEXT and JSPS (Japan);
ARC and DEST (Australia); NSFC (China); DST (India); MOEHRD, KOSEF and
KRF (Korea); KBN (Poland); MES and RFAAE (Russia); ARRS (Slovenia);
SNSF (Switzerland); NSC and MOE (Taiwan); and DOE (USA).


\begin{thebibliography} {99}
\bibitem{foot} Charge-conjugate modes are included throughout this
  paper.
\end{thebibliography}
\end{document}